\documentclass{article}
\usepackage[utf8]{inputenc}

\usepackage{arxiv}

\usepackage[utf8]{inputenc} 
\usepackage[T1]{fontenc}    
\usepackage{hyperref}       
\usepackage{url}            
\usepackage{booktabs}       
\usepackage{amsfonts}       
\usepackage{nicefrac}       
\usepackage{microtype}      
\usepackage{lipsum}		
\usepackage{graphicx}
\usepackage{subcaption} 
\usepackage[numbers]{natbib}
\usepackage{doi}
\usepackage{float} 
\usepackage{dirtytalk}
\usepackage{autonum}
\usepackage{comment}
\usepackage[linesnumbered,ruled,vlined]{algorithm2e}

\input{qcircuit.sty}

\usepackage{acro}

\SetCommentSty{mycommfont}

\SetKwInput{KwInput}{Input}                
\SetKwInput{KwOutput}{Output}  

\usepackage{amsmath,amssymb}  
\usepackage{physics}
\usepackage{bm}

\usepackage[colorinlistoftodos]{todonotes}

\title{Utilizing small quantum computers for machine learning and ground state energy approximation}

\author{
  Stian Bilek \\
  Department of Physics \\
  University of Oslo \\
  Oslo, Norway\\
  \texttt{stian.bilek@fys.uio.no} \\
  \texttt{stianbilek@gmail.com}
}

\hypersetup{
pdftitle={A template for the arxiv style},
pdfsubject={q-bio.NC, q-bio.QM},
pdfauthor={David S.~Hippocampus, Elias D.~Striatum},
pdfkeywords={First keyword, Second keyword, More},
}

\begin{document}

\maketitle

\begin{abstract}
	Quantum circuit partitioning (QCP) is a hybrid quantum-classical approach that aims to simulate large quantum systems on smaller quantum computers. A quantum computation is divided into smaller subsystems and results of measurements on these subsystems are combined using classical processing. In this paper, we propose a QCP strategy to measure an observable on a large quantum system by utilizing several quantum systems of smaller size. The method can be applied to both machine learning and variational ground state energy approximation, and we show that the required calculations and the variance of the gradients can be tailored to scale efficiently with the total number of qubits. Thus it can be utilized to mitigate the well-known problem of barren plateaus. Additionally, the method can be realized by performing simple measurements of Pauli-strings on the separate subsystems, and the gradients can be estimated with common methods such as the parameter-shift rule. We demonstrate the method by approximating the ground state energy of the 1D transverse-field Ising model with periodic boundary conditions, and by classifying handwritten digits. For the ground state energy approximation, we achieved a relative error within the order of 0.1\% for all the tested system sizes. When applied to the classification between the digits 3 and 6, we were able to generalize to out-of-sample data with 100\% accuracy.
\end{abstract}

\tableofcontents

\section{Introduction}

Quantum computers have the potential to solve certain problems that are intractable for classical computers by exploiting the unique properties of quantum mechanics \cite{DeutschJozsa} \cite{ShorPrime}. Consequently, there has been an increasing interest in employing quantum computers for machine learning and ground state approximation tasks to explore whether they can provide a competitive edge in these domains. In the case of machine learning, several quantum algorithms such as Quantum Support Vector Machines \cite{quantumsupportvectormachine}, Quantum Principal Component Analysis \cite{qpca}, Quantum Nearest-Neighbor Classification \cite{knn} and Quantum Principal Component Analysis \cite{qpca} offer speed-ups over their classical analogue under certain assumptions. For ground state approximations, the quantum phase estimation algorithm \cite{kitaev1995quantummeasurementsabelianstabilizer} could potentially be used to efficiently find the eigenvalues of a quantum system \cite{Nielsen_Chuang_2010}.

However, the current noisy intermediate scale quantum (NISQ) era of quantum computing poses several challenges for the practical implementation of quantum algorithms \cite{Preskill2018quantumcomputingin}. We have limitations in the decoherence-times, the number of available qubits, the accuracy of quantum gates and also the connectivity between qubits. Practical demonstrations of quantum speed-ups in the NISQ era are thus reliant on algorithms that respect these hardware limitations. In this regard, new and promising methods specifically tailored for NISQ devices have been developed, such as the employment of Parameterized Quantum Circuits (PQC) \cite{Benedetti_2019} for machine learning and the Variational Quantum Eigensolver (VQE)\cite{VQEPeruzzoProposal} for ground state energy approximations.

Even with functioning algorithms specifically tailored for the NISQ era, the number of qubits one has available puts a hard limit on the space of applicable systems and data sets. For example, a central part of the PQC method is the way which the data set is encoded into the quantum computer. While classical computers can handle high-dimensional data sets and complex models, the relatively short coherence times of NISQ devices restricts the possible ways for which data sets can be encoded reliably. A currently popular encoding scheme is thus to embed each feature of the data set into a respective qubit with the application of single qubit rotations \cite{Benedetti_2019}. While the circuit depth of this approach scales as $\mathcal{O}(1)$, the qubit requirement scales linearly in the number of data set features $p$, thus limiting the data sets for which this can be done reliably. For ground state approximations, the number of qubits required is often decided by the size of the quantum system of interest. 

To address these limitation, one promising approach is to use a divide and conquer strategy, which involves breaking down a large problem into smaller, more manageable subproblems that can be solved independently and then combined to obtain the overall solution. This strategy has been widely used in classical computing for tasks such as sorting \cite{divideconquersorting}, geometric intersection problems \cite{geometricintersection}, and matrix multiplication \cite{STRASSEN1969}. In the context of quantum computing, Quantum circuit partitioning (QCP) is a hybrid quantum-classical approach with the goal of utilizing small quantum computers to simulate larger quantum systems. The idea is to partition large quantum circuits into smaller parts that can be run on a smaller quantum computer than the original circuit. These smaller circuits are measured, before combining the measurement results using classical resources to emulate the full circuit. Several studies have explored this task from different angles. \citeauthor{PhysRevX.6.021043} showed that any $(n+k)$-qubit circuit where each qubit participates in at most $d$ two-qubit gates ($d$-sparse circuits), can be simulated on an $n$-qubit device with classical processing in $2^{\mathcal{O}(kd)}\text{poly}(n)$ steps \cite{PhysRevX.6.021043}. Due to the exponential scaling in $k$ and $d$, this methods practicality is arguably dependent on these parameters being small. \citeauthor{PRXQuantum.3.010346} utilized a similar approach applied to VQE in order to estimate the ground state energy of one-dimensionally coupled Heisenberg antiferromagnetic models on kagome lattices as well as two-dimensional Heisenberg antiferromagnetic models on square lattices \cite{PRXQuantum.3.010346}. Their method is reliant on weak-interactions between circuit partitions, along with subroutines such as the estimation of inner products between quantum states and the Gram-Schmidt procedure. In the context of quantum machine learning, QCP has been applied to digit recognition \cite{Marshall2023highdimensional}, where they successfully utilized quantum computers much smaller than the data dimension. This method is also reliant on the estimation of inner products between quantum states. For a specific family of quantum circuits and observables, QCP has also been shown to mitigate the problem of barren plateaus \cite{QCPBarrenPlateaus}, a current obstacle for many practical quantum computing algorithms. However, the specific family of quantum models they consider does not allow for observables that measure qubits in two partitions simultaneously. 

In this study, we will present a QCP method which aims at mitigating some of the potential drawbacks of present methods. It allows for a subexponential scaling even as the target system size increases, it can be realized by performing simple measurements of Pauli-strings, and it allows for interactions or measurements that act across circuit partitions. The problem of barren plateus can also be mitigated with our approach. Moreover, it can be applied both for machine learning and variational quantum eigensolver tasks. However, the mitigation of these drawbacks does come at the cost of a limitation on the applicable quantum circuits. To illustrate the method, it will be applied for the binary classification task of differentiating between two handwritten digits of the Digit dataset \cite{misc_pen-based_recognition_of_handwritten_digits_81}, as well as for approximating the ground state energy of the 1D transverse-field Ising model with periodic boundary conditions.

The study is structured as follows: Section \ref{sec:Theory} details the theory behind the approach and explains how it can be utilized for both ground state approximation and machine learning tasks. In addition, we show in the final subsection that the problem of barren plateaus can be mitigated. In section \ref{sec:Method} we detail the specific problems we apply our method on, and show how the method is set up.
In section \ref{sec:Results}, we present and discuss the results of our simulations, while we finally conclude and discuss future work in section \ref{sec:Conclusion}. Additional derivations are given in appendix \ref{app:Introduction}.

\section{Theory}
\label{sec:Theory}

\subsection{Quantum states and observables}
\label{subsec:QCPProblemDescription}

Before discussing the specific method, it will help to define what is meant by local operators. We say that an operator is local to a specific subset of the qubits if it acts non-trivially only on said subset, but trivially on all qubits outside of the subset. A trivial operation to a qubit is simply the identity operation $\hat{A}\ket{\psi} = \ket{\psi}$. For example, the four-qubit operator $B = I \otimes X \otimes Y \otimes I + I \otimes I \otimes I \otimes X = I \otimes (X \otimes Y \otimes I + I \otimes I \otimes X)$ only acts trivially on the first qubit, hence it is local to the second, third and fourth qubit.

The aim of the following method is to estimate expectation values of an $n$-qubit state $\ket{\psi}$ with respect to some measurement operator $M$, that is $\bra{\psi} M \ket{\psi}$, by utilizing less than $n$ qubits. We start out with an $n$-qubit system in the ground state and partition these $n$ qubits into $S$ subsystems $\bigotimes_{s=1}^S \ket{0}_s$. The number of qubits in each subsystem, denoted $n_s$, is arbitrary as long as $\sum_{s=1}^S n_s = n$. The next step is to apply a tensor product of unitary operations local to each partition, which allows us to express our quantum state as $\bigotimes_{s=1}^S U_s \ket{0}_s.$ Finally, we apply an arbitrary global unitary $V$ to the full system resulting in the state,
\begin{equation}
    \label{eq:QCPHypothesisState}
    \ket{\psi} = V\bigotimes_{s=1}^S U_s \ket{0}_s.
\end{equation}
The measurement operators we will consider are Pauli-strings of the form $M_m = \sigma^m_i \otimes \sigma^m_j \otimes \cdots \otimes \sigma^m_k$, with $\sigma_i^m \in \{I, \sigma_x, \sigma_y, \sigma_z \}$, but our results can be easily extended to general operators $M = \sum_m c_m M_m$, where $c_m$ are complex coefficients. This is possible due to the linearity of expectation values with respect to a sum of operators. We are thus interested in the quantity, 
\begin{equation}
    \label{eq:QCPHypothesisClass}
    \bra{\psi} M_m \ket{\psi} = [\bigotimes_{s=1}^S \bra{0}_s  U^\dagger_s]  V^\dagger M_m  V \bigotimes_{s=1}^S U_s \ket{0}_s. 
\end{equation}
We can rewrite this expectation value by expressing the global unitary operator as $V = \sum_{l} \lambda_l \sigma^l_p \otimes \sigma^l_q \otimes \cdots \otimes \sigma^l_r$, as long as the coefficients $\lambda_l$ are chosen to ensure unitarity. We have,
\begin{align}
    \bra{\psi} M_m \ket{\psi} &= \sum_{l} \sum_{l'} \lambda_l^* \lambda_{l'} \notag \\
    &[\bigotimes_{s=1}^S \bra{0}_s  U^\dagger_s][\sigma^l_p \otimes \sigma^l_q \otimes \cdots \otimes \sigma_r^l] [\sigma^m_i \otimes \sigma^m_j \otimes \cdots \otimes \sigma^m_k] [\sigma^{l'}_{p} \otimes \sigma^{l'}_{q} \otimes \cdots \otimes \sigma_{r}^{l'}] \bigotimes_{s=1}^S U_s \ket{0}_s \\
    &=\sum_{l} \sum_{l'} \lambda_l^* \lambda_{l'} [\bigotimes_{s=1}^S \bra{0}_s  U^\dagger_s][\sigma^l_p \sigma^m_i \sigma^{l'}_{p} \otimes \sigma^l_q \sigma_j^m \sigma_{q}^{l'} \otimes \cdots \otimes \sigma_r^l \sigma^m_k \sigma^{l'}_{r}]  \bigotimes_{s=1}^S U_s \ket{0}_s \label{eq:firstline} \\
    &= \sum_p d_p [\bigotimes_{s=1}^S \bra{0}_s  U^\dagger_s][\sigma^p_i \otimes \sigma^p_j \otimes \cdots \otimes \sigma^p_k]  \bigotimes_{s=1}^S U_s \ket{0}_s  \label{eq:secondline} \\
    &= \sum_p d_p \prod_{s=1}^S \bra{0}_s U^\dagger_s W^p_s U_s \ket{0}_s \label{eq:thirdline} ,
\end{align}
where $d_p$ are complex coefficients and $W^p_s$ are tensor products of Pauli-operators local to subset $s$. Going from Eq. \ref{eq:firstline} to Eq. \ref{eq:secondline}, we utilized the fact that a product of Pauli-operators can be written as $\sigma_a\sigma_b\sigma_c = e^{i\alpha} \sigma_d$, allowing us to re-express the observable in Eq. \ref{eq:firstline} as $\sum_{l}\sum_{l'}\lambda_l^* \lambda_l [\sigma^l_p \sigma^m_i \sigma^{l'}_{p} \otimes \sigma^l_q \sigma_j^m \sigma_{q}^{l'} \otimes \cdots \otimes \sigma_r^l \sigma^m_k \sigma^{l'}_{r}] = d_p [\sigma^p_i \otimes \sigma^p_j \otimes \cdots \otimes \sigma^p_k] = \sum_p d_p \bigotimes_{s=1}^S W_s^p$. The key takeaway from Eq. \ref{eq:thirdline} is that we are able to express the expectation value on the full $n$-qubit system in Eq. \ref{eq:QCPHypothesisClass} in terms of products of measurements on the $S$ subsystems. Even so, the practicality of the method relies on the scaling of the number of expectation values and complex coefficients in the number of qubits. We will discuss this along with how we can apply Eq. \ref{eq:thirdline} for machine learning tasks and VQE tasks in the upcoming sections.

\subsection{Supervised learning}
\label{subsec:QCPSupervisedLearning}
In this section, we will show how the expectation value in Eq. \ref{eq:QCPHypothesisClass} can be utilized for supervised learning tasks on quantum computers with less qubits than in the original quantum state. Assume that a sample of our dataset features is given by $\vec{x} \in \mathrm{R}^p$. A common way to encode these $p$ features into a quantum state is to embed each feature into a respective qubit with the application of single qubit rotations \cite{Benedetti_2019}. This task requires a quantum computer of at least $p$ qubits. If only an $m < p$-qubit quantum computer is available, we divide the features into $S$ subsets of at most size $m$, denoted $\vec{x}^{(s)}$ for $s = 1,2,\dots,S$.

First, we rewrite the quantum state in Eq. \ref{eq:QCPHypothesisState} as
\begin{equation}
    \label{eq:ParameterizedAnsatz}
    \ket{\psi} = V \bigotimes_{s=1}^S U_s(\vec{x}^{(s)}, \theta^{(s)}_1, \theta^{(s)}_2, \dots) \ket{0}_s ,
\end{equation}
where $\theta^{(s)}_i$ is the $i$'th rotational parameter for subset $s$. For many machine learning tasks, the models predictions does not need to associate with the measurement of an hermitian operator $M_m$ on a normalized quantum state $V \bigotimes_{s=1}^S U_s \ket{0}_s$. This means that the observable $\sum_p d_p \bigotimes_{s=1}^S W_s^p$ in Eq. \ref{eq:thirdline} can be chosen arbitrarily, for example as a linear combination of some $n$-qubit Pauli strings. Given $\sum_p d_p \bigotimes_{s=1}^S W_s^p$ of such form, we use Eq. \ref{eq:thirdline} and define our quantum model as
\begin{align}
    f(\vec{x}, \theta^{(1)}_1, \theta^{(1)}_2, &\dots, \theta^{(2)}_1, \theta^{(2)}_2, \dots, d_1, d_2, \dots) = \notag \\
    & \sum_p d_p \prod_{s=1}^S \bra{0}_s U^\dagger_s(\vec{x}^{(s)}, \theta^{(s)}_1, \theta^{(s)}_2, \dots) W^{p}_s U_s(\vec{x}^{(s)}, \theta^{(s)}_1, \theta^{(s)}_2, \dots) \ket{0}_s. \label{eq:QPCHypothesisClassSupervisedLearning}
\end{align}
This expression, or some classical processing step, $g[f(\vec{x}, \theta^{(1)}_1, \theta^{(1)}_2, \dots, \theta^{(2)}_1, \theta^{(2)}_2, \dots, d_1, d_2, \dots)]$, gives us the output of our supervised model.

\subsubsection{Gradients}
\label{subsubsec:SupervisedLearningGradients}
Given a differentiable loss function, training supervised models with gradient descent methods require calculating the partial derivatives of the expression in Eq. \ref{eq:QPCHypothesisClassSupervisedLearning}, wrt. the trainable parameters of our model. For the coefficients, we have
\begin{equation}
    \label{eq:GradientWrtComplexCoeffs}
    \frac{\partial f(\vec{x}, \theta^{(1)}_1, \theta^{(1)}_2, \dots, \theta^{(2)}_1, \theta^{(2)}_2, \dots, d_1, d_2, \dots)}{\partial d_q} = \prod_{s=1}^S \bra{0}_s U^\dagger_s(\vec{x}^{(s)}, \theta^{(s)}_1, \theta^{(s)}_2, \dots) W^q_s U_s(\vec{x}^{(s)}, \theta^{(s)}_1, \theta^{(s)}_2, \dots) \ket{0}_s,
\end{equation}
which is straight forward to evaluate.
For the rotational parameters, we have
\begin{align}
    &\frac{\partial f(\vec{x}, \theta^{(1)}_1, \theta^{(1)}_2, \dots, \theta^{(2)}_1, \theta^{(2)}_2, \dots, d_1, d_2, \dots)}{\partial \theta^{(q)}_i} = \notag \\
    &\sum_p d_p \frac{\partial}{\partial \theta^{(q)}_i}\left(\bra{0}_q U^\dagger_q(\vec{x}^{(q)}, \theta^{(q)}_1, \theta^{(q)}_2, \dots) W^{p}_q U_q(\vec{x}^{(q)}, \theta^{(q)}_1, \theta^{(q)}_2, \dots) \ket{0}_q \right) \notag \\
    & \cdot \prod_{s\neq q}^S \bra{0}_s U^\dagger_s(\vec{x}^{(s)}, \theta^{(s)}_1, \theta^{(s)}_2, \dots) W^{p}_s U_s(\vec{x}^{(s)}, \theta^{(s)}_1, \theta^{(s)}_2, \dots) \ket{0}_s, \label{eq:QPCGradientWrtRotationalParams}
\end{align}
where 
\begin{align}
    \frac{\partial}{\partial \theta^{(q)}_i}\left(\bra{0}_q U^\dagger_q(\vec{x}^{(q)}, \theta^{(q)}_1, \theta^{(q)}_2, \dots) W^p_q U_q(\vec{x}^{(q)}, \theta^{(q)}_1, \theta^{(q)}_2, \dots) \ket{0}_q \right)
    \label{eq:Gradientofsubsystem}
\end{align}
can be evaluated with the parameter-shift rule \cite{ParameterShiftRule}. 

\subsubsection{Computational complexities}
\label{subsec:QCPComputationalComplexities}
As mentioned earlier, the observable $\sum_p d_p \bigotimes_{s=1}^S W_s^p$ can be chosen arbitrarily for many machine learning tasks. If chosen as a linear combination of all possible $n$-qubit Pauli strings, the maximum number of expectation values for subsystem $s$ is given by the number of unique combinations of the Pauli-basis, $\{I, X, Y, Z\}$, on $n_s$ qubits, that is $4^{n_s}$. The total number of expectation values is therefore at most $\sum_{s=1}^S 4^{n_s}$. One can show that this number can be made polynomial in the number of qubits $n$. If we partition the system into $S = n/n_s$ equally sized subsystems of $n_s = \beta\log_2(n)$ qubits, for some constant $\beta$, the total number of expectation values is given by
\begin{equation}
    \label{eq:QCPPolynomialNumberOfExpectationValues}
    n_E \leq S \cdot 4^{n_s} = \frac{n}{\beta\log_2(n)}2^{2\beta \log_2(n)} = \frac{n^{2\beta+1}}{\beta\log_2 (n)}.
\end{equation}
Despite the polynomial scaling of the number of expectation values, we still require one coefficient $d_p$ for every possible $n$-qubit combination of the Pauli-basis. The scaling in the number of coefficients is hence $4^n$. However, one could evaluate all the required expectation values in polynomial time, and then make an informed choice of a number of terms from Eq. \ref{eq:QPCHypothesisClassSupervisedLearning} polynomial in $n$. One approach is to evaluate the gradients in Eq. \ref{eq:Gradientofsubsystem} and chose to form $\sum_p d_p \bigotimes_{s=1}^S W_s^p$ as a combination of the Pauli-strings $W_s^p$ which provide the gradient with the largest magnitude. Another approach can be seen by inspecting Eq. \ref{eq:GradientWrtComplexCoeffs}. We see from this equation that the rate of change of the expectation value wrt. the coefficients is proportional to products of expectation values on each subsystem. Further, for a given loss function $L(g[f(\vec{\phi})])$, the gradient wrt. the parameter $\phi_i$ is given by
\begin{equation}
\label{eq:QCPLossGeneral}
\frac{\partial L(g[f(\vec{\phi})])}{\partial \phi_i} = \frac{\partial L(g[f(\vec{\phi})])}{\partial g[f(\vec{\phi})]} \frac{\partial g[f(\vec{\phi})]}{\partial f(\vec{\phi})} \frac{\partial f(\vec{\phi})}{\partial \phi_i}.
\end{equation}
When $f(\vec{\phi})$ is the expectation value in Eq. \ref{eq:QPCHypothesisClassSupervisedLearning}, it follows that products of expectation values which are close to zero will lead to a small factor in the rate of change of the loss function wrt. our coefficients. We thus evaluate all possible expectation values on all subsystems, and construct $\sum_p d_p \bigotimes_{s=1}^S W_s^p$ by combining operators $W^p_s$ which resulted in expectation values with the largest magnitudes. If a polynomial in $n$ such combinations are chosen, it follows that the number of coefficients is also polynomial in $n$.

We can also limit the number of terms in $\sum_p d_p \bigotimes_{s=1}^S W_s^p$ by considering it as a linear combination of up to $p$-qubit operators, where $p<n$. The number of ways to chose $p$ qubits from $n$ qubits is $\binom{n}{p}$, and there are $4^{p}$ ways to combine the Pauli-basis for each of these combinations. Thus the number of coefficients required is $\sum_{k=1}^{p} \binom{n}{k}4^k$. We can utilize the upper bound $\binom{n}{k} \leq (\frac{en}{k})^k$ \cite{10.5555/233228.233232} to rewrite this as $n_T \leq \sum_{k=1}^{p} (\frac{en}{k})^k4^k$. For fixed $p$ and large $n$, the term with the highest power, $(\frac{en}{p})^{p} 4^{p}$,  grows the fastest. This results in an asymptotical complexity of $\mathcal{O}(n^{p})$.

\subsection{Variational quantum eigensolvers}
\label{subsec:VQE}
Given a Hamiltonian $H = \sum_m \lambda_m M_m$, where $\lambda_i$ are real coefficients and $M_m$ are Pauli-strings, the aim with VQE is to minimize the energy,
\begin{equation}
    \label{eq:EnergyExpression}
    E = \bra{\psi(\vec{\phi})} H \ket{\psi(\vec{\phi})},
\end{equation}
by varying the parameters $\vec{\phi}$. Instead of measuring $H$ directly on the state $\ket{\psi} = V \bigotimes_{s=1}^S U_s(\vec{\theta}_s) \ket{0}_s$, our method relies on measurements of the products $V^\dagger H V$ on the state $\bigotimes_{s=1}^S U_s(\vec{\theta}_s) \ket{0}_s$. For $V = \sum_i c_i P_i$, we can according to Eq. \ref{eq:QCPHypothesisClass} write out the energy expression as 
\begin{align}
    E(\theta^{(1)}_1, \theta^{(1)}_2, \dots, \theta^{(2)}_1, \theta^{(2)}_2, \dots, c_1, c_2, \dots) &= \left[\bigotimes_{s=1}^S\bra{0}_s U^\dagger_s(\vec{\theta}_s)\right] V^\dagger H V \left[ \bigotimes_{s=1}^S U_s(\vec{\theta}_s) \ket{0}_s \right] \notag \\ 
    &= \sum_p d_p(c_1,c_2,\dots, \lambda_1, \lambda_2, \dots) \prod_{s=1}^S \bra{0}_s U^\dagger_s(\vec{\theta}_s) W^p_s U_s(\vec{\theta}_s) \ket{0}_s. 
    \label{eq:EnergyExpressionWrittenOut}
\end{align}
In order to utilize our approach for methods such as the variational quantum eigensolver (VQE), one is reliant on the unitarity of the operator $V$. Thus, the measurement of $V^\dagger M_m V = \sum_p d_p \bigotimes_{s=1}^S W_s^p$ will need to correspond to the measurement of the hermitian operator $M_m$ on a normalized state $V \bigotimes_{s=1}^S U_s\ket{0}_s$. Since $V$ is of dimensionality $2^n \otimes 2^n$, efficient construction of this unitary and calculation of the products $V^\dagger M_m V$ is not straight forward. However, by restricting the form of $V$ this can be achieved efficiently. We first define a set of Pauli-strings $\{P_1, P_2, \dots, P_N \}$ and real scalars $\{c_1, c_2, \dots, c_N\}$. We then express $V$ as,
\begin{align}
    V &= \prod_{i=1}^N [\cos (c_i)I + i\sin (c_i)P_i] \notag \\
    &= \prod_{i=1}^N \overline{P}_i.
    \label{eq:Vexpression}
\end{align}

This ensures the unitarity of $V$ since 
\begin{align}
    \overline{P}_i^\dagger \overline{P}_i &= [\cos (c_i)I - i\sin (c_i)P_i][\cos (c_i)I + i\sin (c_i)P_i] \notag \\
    &= \cos^2(c_i)I + \sin^2(c_i)I = I.
\end{align}
We will return to detailing the construction of the Pauli-strings and the scaling of Eq. \ref{eq:Vexpression} in section \ref{subsubsec:VQEComputationComplexities}.

\subsubsection{Gradients}
\label{subsubsecVQEGradients}
We can minimize the energy expression in Eq. \ref{eq:EnergyExpressionWrittenOut} by finding its gradient wrt. the rotational parameters, $\vec{\theta}_s$ and the coefficients, $c_i$. The rotational parameters follow a similar expression as for the supervised learning case, that is,
\begin{align}
    &\frac{\partial E(\theta^{(1)}_1, \theta^{(1)}_2, \dots, \theta^{(2)}_1, \theta^{(2)}_2, \dots, c_1, c_2, \dots)}{\partial \theta^{(q)}_i} = \notag \\
    &\sum_p d_p \frac{\partial}{\partial \theta^{(q)}_i}\left(\bra{0}_q U^\dagger_q( \theta^{(q)}_1, \theta^{(q)}_2, \dots) W^{p}_q U_q(\theta^{(q)}_1, \theta^{(q)}_2, \dots) \ket{0}_q \right) \notag \\
    & \cdot \prod_{s\neq q}^S \bra{0}_s U^\dagger_s( \theta^{(s)}_1, \theta^{(s)}_2, \dots) W^{p}_s U_s( \theta^{(s)}_1, \theta^{(s)}_2, \dots) \ket{0}_s, \label{eq:VQEGradientWrtRotationalParams}
\end{align}
where 
$$\frac{\partial}{\partial \theta^{(q)}_i}\left(\bra{0}_q U^\dagger_q( \theta^{(q)}_1, \theta^{(q)}_2, \dots) W^p_q U_q(\theta^{(q)}_1, \theta^{(q)}_2, \dots) \ket{0}_q \right)$$
can be evaluated with the parameter-shift rule \cite{ParameterShiftRule}.
For the coefficients we have,
\begin{align}
    &\frac{\partial E(\theta^{(1)}_1, \theta^{(1)}_2, \dots, \theta^{(2)}_1, \theta^{(2)}_2, \dots, c_1, c_2, \dots)}{\partial c_i} = \notag \\
    &\sum_p \frac{\partial d_p(c_1,c_2,\dots)}{\partial c_i} \prod_{s=1}^S \bra{0}_s U^\dagger_s(\vec{\theta}_s) W^p_s U_s(\vec{\theta}_s) \ket{0}_s,
    \label{eq:VQEGradientWrtCoefficients}
\end{align}
where the products $\prod_{s=1}^S \bra{0}_s U^\dagger_s(\vec{\theta}_s) W^p_s U_s(\vec{\theta}_s) \ket{0}_s$ are already evaluated when producing the output.

\subsubsection{Computational complexities and the choice of Pauli-strings}
\label{subsubsec:VQEComputationComplexities}
Expressing $V$ as in Eq. \ref{eq:Vexpression} and expanding the product results in a sum of $2^N$ terms. The observable $V^\dagger M_m V$ will thus consist of $2^{2N}$ terms, and if $N = \beta \log_2 (n)$, the total number of terms is $n^{2\beta}$. If a logarithmic in $n$ number of Pauli-strings is not sufficient, we can restrict the choice of $\{P_1, P_2, \cdots, P_N\}$ in order to achieve a scaling independent of $N$.

We construct a set of Pauli strings \(\{P_1, P_2, \cdots, P_N\}\) with specific non-commutation properties to achieve scaling independent of $N$, and we start with the assumption that \(P_{i_1}\) is the first Pauli string which is non-commutative with \(M_m\). The target observable \(V^\dagger M_m V\) can then be written as

\begin{align}
    V^\dagger M_m V &= \overline{P_N}^\dagger \cdots \overline{P_{i_1}}^\dagger \cdots \overline{P_1}^\dagger M_m \overline{P_1} \cdots \overline{P_{i_1}} \cdots \overline{P_N} \notag \\
    &= \overline{P_N}^\dagger \cdots \overline{P_{i_1}}^\dagger M_m \overline{P_{i_1}} \cdots \overline{P_N}.
\end{align}

Next, we identify the Pauli string \(P_{i_2}\), which is the next Pauli string that is non-commutative with either \(P_{i_1}\) or \(M_m\). We then have

\begin{align}
    V^\dagger M_m V &= \overline{P}_N^\dagger \cdots \overline{P}_{i_2}^\dagger \cdots \overline{P}^\dagger_{i_1 + 1} \overline{P}^\dagger_{i_1} M_m \overline{P}_{i_1} \overline{P}_{i_1 + 1} \cdots \overline{P}_{i_2} \cdots \overline{P}_N \notag \\
    &= \overline{P}_N^\dagger \cdots \overline{P}_{i_2}^\dagger \overline{P}_{i_1}^\dagger M_m \overline{P}_{i_1} \overline{P}_{i_2} \cdots \overline{P}_N.
\end{align}

We can proceed similarly by finding the next Pauli string $P_{i_3}$, which is non-commutative with either $M_m$, $P_{i_1}$, or $P_{i_2}$, and then continue this process for the next Pauli string $P_{i_4}$. Assuming there exists a total of $K$ such Pauli strings for $M_m$, we get

\begin{align}
    V^\dagger M_m V &= \overline{P_{i_K}}^\dagger \cdots \overline{P_{i_2}}^\dagger \overline{P_{i_1}}^\dagger M_m \overline{P}_{i_1} \overline{P_{i_2}} \cdots \overline{P_{i_K}}.
    \label{eq:VdaggMVspecificproperties}
\end{align}

This results in $2^{2K}$ expectation values per observable $M_m$, and is thus constant in $n$ for fixed $K$. We can also allow $K$ to scale with $n$ by chosing $K = \beta \log_2(n)$, resulting in $n^{2\beta}$ terms per observable $M_m$. 

Given a Hamiltonian of $L$ terms, $H = \sum_{m=1}^L \lambda_m M_m$, the product $V^\dagger H V$ can be expanded to yield the new observable $D = V^\dagger H V = \sum_{p}d_p \bigotimes_{s=1}^S W_s^p$. By following the strategy outlined above, this observable will contain at most $L2^{2K}$ Pauli strings. The number of expectation values that has to be evaluated depends on $H$ and the choice of the Pauli strings $P_i$. Worst case scenario is that every term in $D$ acts on all the $n$ qubits, requiring one expectation value per each of the $S$ subsystems, per term. This results in $SL2^{2K}$ expectation values, which can be made subexponential in $n$ as long as the Hamiltonian can be expressed with a number of terms $L$ subexponential in $n$.

\subsection{Barren plateaus}
\label{sec:barrenplateaus}
\citeauthor{VanishingGradients} showed that for a wide range of parameterized circuits, the probability that a gradient along any direction being non-zero vanishes exponentially in the number of qubits \cite{VanishingGradients}. This results in a significant challenge in the training of such quantum models, hence it is natural to explore the behaviour of the gradients for our method. We are thus interested in the variance of the gradients in Eqs. \ref{eq:QPCGradientWrtRotationalParams} and \ref{eq:VQEGradientWrtRotationalParams}, and how these scale with $n$. We can write the variance of both of these gradients as
\begin{align}
    \text{Var}[&\frac{\partial f(\theta^{(1)}_1, \theta^{(1)}_2, \dots, \theta^{(2)}_1, \theta^{(2)}_2, \dots, d^{(1)}_1, \dots)}{\partial \theta^{(q)}_i}] = \notag \\
    &\text{Var}[\sum_p d_p \frac{\partial}{\partial \theta^{(q)}_i}\left(\bra{0}_q U^\dagger_q( \theta^{(q)}_1, \theta^{(q)}_2, \dots) W^{p}_q U_q(\theta^{(q)}_1, \theta^{(q)}_2, \dots) \ket{0}_q \right) \notag \\
    & \cdot \prod_{s\neq q}^S \bra{0}_s U^\dagger_s( \theta^{(s)}_1, \theta^{(s)}_2, \dots) W^{p}_s U_s( \theta^{(s)}_1, \theta^{(s)}_2, \dots) \ket{0}_s] \notag \\
    &= \text{Var}[\sum_p d_p R_q^{(p)} \prod_{s\neq q} B_s^{(p)}],
    \label{eq:QPCVarGradientWrtRotationalParams}
\end{align}
where 
\begin{align}
    R^{(p)}_q = \frac{\partial}{\partial \theta^{(q)}_i}\left(\bra{0}_q U^\dagger_q(\theta^{(q)}_1, \theta^{(q)}_2, \dots) W^{p}_q U_q(\theta^{(q)}_1, \theta^{(q)}_2, \dots) \ket{0}_q \right),
\end{align}
and
$$B_s^{(p)} = \bra{0}_s U^\dagger_s(\theta^{(s)}_1, \theta^{(s)}_2, \dots) W^{p}_s U_s(\theta^{(s)}_1, \theta^{(s)}_2, \dots) \ket{0}_s.$$ 
We omitted the inclusion of $\vec{x}^{(s)}$ since we can consider these implicitly included in $U_s$ for the supervised learning case. We continue by utilizing that, 
\begin{align}
    \text{Var}[\sum_i X_i] = \sum_i \text{Var}[X_i] + 2 \sum_{i<j}\text{Cov}[X_i,X_j],
\end{align}
which gives,
\begin{align}
    \text{Var}[\sum_i d_i R_q^{(i)} \prod_{s\neq q} B_s^{(i)}] &= \sum_i \text{Var}[d_i R_q^{(i)} \prod_{s\neq q} B_s^{(i)}] + 2 \sum_{i<j}\text{Cov}[d_i R_q^{(i)} \prod_{s\neq q} B_s^{(i)},d_j R_q^{(j)} \prod_{s\neq q} B_s^{(j)}].
\end{align}
First we deal with the covariance by utilizing $\text{Cov}(X,Y) = \text{E}[XY] - \text{E}[X]\text{E}[Y]$. By assuming that each random circuit, $B^{(i)}_s$, resembles a unitary 2-design, we utilize that the expected value of the gradients $\text{E}[R^{(p)}_q] = 0$ \cite{VanishingGradients}. This assumption also leads to $\text{E}[B_s^{(i)} B_s^{(j)}] = 0$, which is shown in Appendix \ref{app:BiBj}. Further,
since $R^{(p)}_q$ and $R^{(p)}_{q'}$ for $q\neq q'$ are measurements on different subsystems, we consider $R^{(p)}_q$ to be independent of all variables except $R^{(p')}_q$ for all $p'$. The same can be argued for $B_s^{(p)}$, which gives,
\begin{align}
    \text{Cov}[d_i R_q^{(i)} \prod_{s\neq q} B_s^{(i)},d_j R_q^{(j)} \prod_{s\neq q} B_s^{(j)}] &= \text{E}[d_i R_q^{(i)} (\prod_{s\neq q} B_s^{(i)}) d_j^* R_q^{(j)} \prod_{s\neq q} B_s^{(j)}] \notag \\
    &- \text{E}[d_i R_q^{(i)} \prod_{s\neq q} B_s^{(i)}]\text{E}[d_j^* R_q^{(j)} \prod_{s\neq q} B_s^{(j)}] \notag \\
    &= \text{E}[d_i d_j^*]\text{E}[R_q^{(i)}R_q^{(j)}] \prod_{s\neq q} \text{E}[B_s^{(i)} B_s^{(j)}] \notag \\
    &- \text{E}[d_i]\text{E}[d_j^*]\text{E}[R_q^{(i)}]\text{E}[R_q^{(j)}] \text{E}[(\prod_{s\neq q} B_s^{(i)})]\text{E}[\prod_{s\neq q} B_s^{(j)}] \notag \\
    &= 0,
    \label{eq:CovarianceGeneral}
\end{align}
where $\text{E}[R^{(p)}_q] = 0$ and $\text{E}[B_s^{(i)} B_s^{(j)}] = 0$ has been utilized when transitioning to the final line. The variance can then be written as,
\begin{align}
    \text{Var}[\sum_i d_i R_q^{(i)} \prod_{s\neq q} B_s^{(i)}] &= \sum_i \text{Var}[d_i R_q^{(i)} \prod_{s\neq q} B_s^{(i)}]  \notag \\
    &= \sum_i \biggl( \text{E}[d_i R_q^{(i)} (\prod_{s\neq q} B_s^{(i)}) d_i^* R_q^{(i)} (\prod_{s\neq q} B_s^{(i)})] - \text{E}[d_i R_q^{(i)} \prod_{s\neq q} B_s^{(i)}]^2 \biggr) \notag \\
    &= \sum_i \text{E}[d_i d_i^*] \text{E}[(R^{(i)}_q)^2] \prod_{s\neq q} \text{E}[(B_s^{(i)})^2], \label{eq:QCPVarianceNoAssumptions}
\end{align}
by again utilizing that $\text{E}[R^{(i)}_q] = 0$.

In order to find the scaling of the variance in Eq. \ref{eq:QCPVarianceNoAssumptions}, we will need to evaluate how each of the components, $\text{E}[d_i d_i^*]$, $\text{E}[(R^{(i)}_q)^2]$ and $\text{E}[(B_s^{(i)})^2]$ scales in the number of qubits $n$. If all these can be made to scale polynomially in $n$, the same can be said of the products in Eq. \ref{eq:QCPVarianceNoAssumptions}.
Assuming $n_s = \beta \log{n}$ qubits in each subsystem implies that we can construct our ansatzes such that the gradients $R^{(i)}_q$ has zero mean and a variance $\text{E}[(R^{(i)}_q)^2]$ which scales polynomially in the total number of qubits, $n$ \cite{QCPBarrenPlateaus}. In Appendix \ref{app:Bsquared}, we show that terms of $D = \sum_p d_p \bigotimes_{s=1}^S W_s^p$ which are local to a smaller subset of $K$ of the subsystems have the appropriate scaling for $\prod_{s\neq q} \text{E}[(B_s^{(i)})^2]$. What is left is the term $\text{E}[d_i d_i^*]$. This term will depend on the specific use case, so we will discuss supervised learning and VQE separately.

\subsubsection{Supervised learning}
For the supervised learning scenario, we are free to chose the distribution of the coefficients $d_i$ in $D = \sum_p d_p \bigotimes_{s=1}^S W_s^p$. If these are considered normally distributed with zero mean and variance $\sigma^2_d$, we have $E[d_i d_i^*] = \sigma^2_d$. This can be considered constant in $n$. This choice will thus result in the appropriate scaling of the variance.

\subsubsection{VQE}
For the VQE scenario, the term $E[d_i d_i^*]$ depends on the form of $V$. For the form explained in section \ref{subsec:VQE}, we need to find $E[d_i d_i^*]$ wrt. the scalars $c_i$. The coefficients $d_i$ are found by calculating, $$V^\dagger M_m V = \overline{P}^\dagger_N \cdots \overline{P}^\dagger_1 M_m \overline{P}_1 \cdots \overline{P}_N, $$ where,
\begin{align}
    V &= \prod_{i=1}^N [\cos (c_i)I + i\sin (c_i)P_i] \notag \\
    &= \prod_{i=1}^N \overline{P}_i.
\end{align}
We begin by calculating the following part,
$$\overline{P}^\dagger_1 M_m \overline{P}_1 = \cos^2 (c_1)I + i\cos (c_1) \sin (c_1)[M_m, P_1] + \sin^2 (c_1) P_1 M_m P_1,$$
and we observe that continuing calculating the products results in a linear combination of Pauli strings, $\sum_{i=1}^{2^{2N}} d_i O_i$, with coefficients given by,
\begin{align}
    d_i(c_1,c_2,\dots,c_n) = \prod_{i=1}^N f_i(c_i).
    \label{eq:d_coefficients_VQE}
\end{align}
The product in Eq. \ref{eq:d_coefficients_VQE} consists of all possible combinations of $f_i(c_i) \in \{\cos^2(c_i), \pm i \cos(c_i) \sin(c_i), \sin^2(c_i)\}.$
If constructing the operator $V$ in the way specified in section \ref{subsubsec:VQEComputationComplexities}, the number of coefficients $d_i$ and the number of factors in the product of Eq. \ref{eq:d_coefficients_VQE} will instead be $2^{2K}$ and $K$, respectively. Assume that $d_i$ includes $k$ terms of the kind $\cos^{2}(c_i)$, and $l$ terms of the kind $\pm i \cos(c_i) \sin(c_i)$ and $m$ terms of the kind $\sin^2(c_i)$. We can then express $E[d_i d_i^*]$ as
\begin{align}
    E[d_i d_i^*] &= E[\prod_{j=1} f_j(c_j). \prod_{j=1} f^*_j(c_j).] \notag \\
    &= \left(\prod_{a=1}^k E[\cos^4(c_{i_a})]\right) \left( \prod_{a=k+1}^{l+k} E[\cos^2(c_{i_a}) \sin^2(c_{i_a})]\right) \left(\prod_{a=l+k+1}^{m+l+k} E[\sin^4(c_{i_a})]\right).
    \label{eq:VQEdexpectation}
\end{align}
Now if each $c_i$ is drawn from the uniform distribution $[0,2\pi]$, we get the following expectation values for each term, $1/2\pi \int_{0}^{2\pi} \cos^4(c) dc = 3/8$, and $1/2\pi \int_{0}^{2\pi} \cos^2(c)\sin^2(c) dc = 1/8$ and $1/2\pi \int_{0}^{2\pi} \sin^4(c) dc = 3/8$. Inserting into Eq. \ref{eq:VQEdexpectation} gives
\begin{align}
    E[d_i d_i^*] &= \left(\frac{3}{8}\right)^k \left(\frac{1}{8}\right)^l \left(\frac{3}{8}\right)^m \notag \\
    \label{eq:VQEfinalexpectation}
\end{align}
Given $K$ factors in the product in Eq. \ref{eq:d_coefficients_VQE}, the worst-case scenario is when $k=0$, $m=0$ and $l=K$. Considering $K = \beta \log_2 (n)$ then gives
\begin{align}
    \label{eq:VQEExpectationvalueOrder}
    E[d_i d_i^*] = \mathcal{O}(\frac{1}{2^{3K}}) = \mathcal{O}(\frac{1}{n^{3\beta}}),
\end{align}
which scales polynomially in the number of qubits.

\subsubsection{Efficient gradient estimation in the presence of barren plateaus}
It is important to note that although the gradient associated with measurements of Pauli strings acting globally on all subsystems diminishes exponentially with the number of qubits, the gradient is actually computed from the products of measurements on each individual subsystem. As discussed in \cite{VanishingGradients}, to avoid a random walk in parameter space, one requires at least $1/||g||^\alpha$ measurements, where $||g||$ is the norm of the gradient. The norm of the gradient for a single subsystem vanishes exponentially with the number of qubits in that subsystem, but it vanishes polynomially with the total number of qubits if the subsystem's qubit count is logarithmic relative to the total number of qubits. Therefore, we can efficiently estimate the gradient for the entire system, provided that the products of measurements on each separate subsystem are within machine precision. Even though the total gradient is still exponentially vanishing, utilizing momentum-based gradient methods could then help to escape the barren plateau. However, it should be noted that this reasoning neglects the effect of accumulated errors from measurements on each subsystem, which could potentially impact the accuracy of the overall gradient estimate.

\section{Method}
\label{sec:Method}
\subsection{Ground state energy}
\label{subsec:MethodGroundState}
In this section, we will introduce our Hamiltonian, as well as how the we construct our ansatz for the ground state.
\subsubsection{The Hamiltonian}
We will consider the 1D transverse-field Ising Hamiltonian with periodic boundary conditions. The Hamiltonian for $n$ spins is given by,
\begin{align}
    H = -J \sum_{i=1}^n X_i X_{i+1} - h\sum_{i=1}^n Z_i,
    \label{eq:IsingHamiltonian}
\end{align}
where $J$ is the interaction strength between neighbooring spins and $h$ is the strength of the transverse field. The periodic boundary conditions imply $X_{n+1} = X_{1}$.

\subsubsection{Subsystem unitaries}
\begin{figure}[H]
\Qcircuit @C=1.7em @R=1em {
& \gate{R_y(\theta_s^{(1)})} & \ctrl{1} & \qw & \qw & \qw & \qw & \qw & \qw & \gate{R_y(\theta_s^{(n+1)})} & \ctrl{1} & \qw & \qw & \qw & \qw & \qw & \qw & \qw\\
& \gate{R_y(\theta_s^{(2)})} & \targ & \ctrl{1} & \qw & \qw & \qw & \qw & \qw & \gate{R_y(\theta_s^{(n+2)})} & \targ & \ctrl{1} & \qw & \qw & \qw & \qw & \qw & \qw\\
& \gate{R_y(\theta_s^{(3)})} & \qw & \targ & \ctrl{1} & \qw & \qw & \qw & \qw & \gate{R_y(\theta_s^{(n+3)})} & \qw & \targ & \ctrl{1} & \qw & \qw & \qw & \qw & \qw\\
& \gate{R_y(\theta_s^{(4)})} & \qw & \qw & \targ & \ctrl{1} & \qw & \qw & \qw & \gate{R_y(\theta_s^{(n+4)})} & \qw & \qw & \targ & \ctrl{1} & \qw & \qw & \qw & \qw \\
& \vdots & & & \ddots & & \ddots & & & \vdots & & & \ddots & & \ddots & & & \\
& \gate{R_y(\theta_s^{(n)})} & \qw & \qw & \qw & \qw & \qw & \qw & \targ & \gate{R_y(\theta_s^{(2n)})} & \qw & \qw & \qw & \qw & \qw & \qw & \targ & \qw
}
\caption{The parameterized quantum circuit for subsystem $s$. We apply $R_y$ rotations to each qubit with an angle specified by the tunable parameters $\vec{\theta}_s^{(k)}$, followed by CNOT gates to produce entanglement. The procedure is done twice.}
\label{fig:VQEAnsatz}
\end{figure}
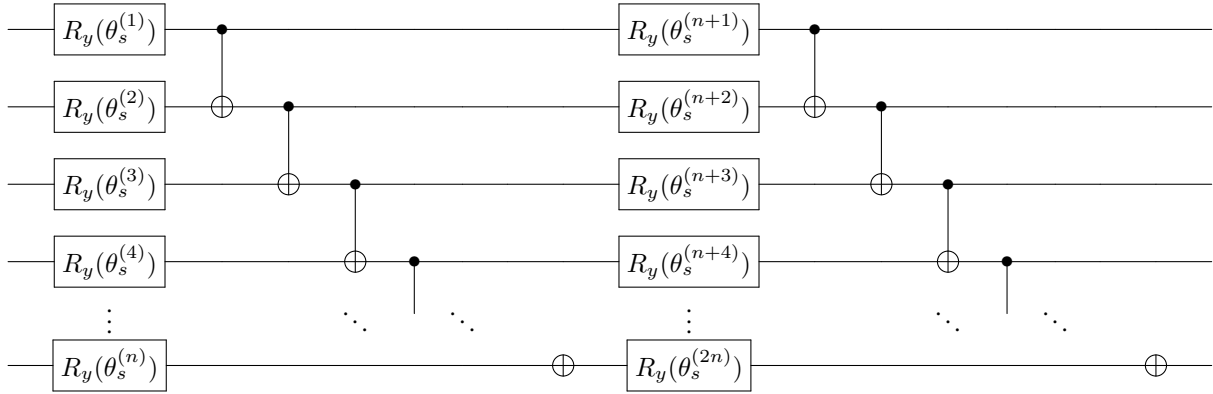

The parameterized quantum circuit (PQC) for each subsystem, that is $U_s(\vec{\theta}_s)$ in Eq. \ref{eq:EnergyExpressionWrittenOut}, is shown in Fig. \ref{fig:VQEAnsatz}.

\subsubsection{Global unitary}
We aim to approximate the ground state energy of Eq. \ref{eq:IsingHamiltonian} for $n=4, 6, 8, 10, 12, 16$ spins by defining the global unitary $V$ according to Eq. \ref{eq:Vexpression}. The Pauli-strings $P_i$ utilized for the construction of $V$, along with the number of subsystems is shown in Tab. \ref{tab:Configuration}.

\begin{table}[h]
\centering
\caption{Table showing the number of spins, $n$, subsystems $S$ and the set of Pauli-strings $\{P_i\}$ utilized for our ground state approximation. We divided the qubits into equally sized subsystems of $n/S$ qubits and the Pauli-strings are chosen to act along and between borders of these subsystems.}
\label{tab:Configuration}
\begin{tabular}{|c|c|l|}
\hline
$n$ & $S$ & $\{P_i\}$ \\ \hline
4   & 2   & $\{Z_2Y_3, Y_1Z_4\}$ \\ \hline
6   & 2   & $\{Z_3Y_4, Y_1Z_6\}$ \\ \hline
8   & 2   & $\{Z_3Y_4, Z_4Y_5, Z_5Y_6, Z_7Y_8, Y_1Z_8, Z_1Y_2\}$ \\ \hline
10  & 2   & $\{Z_4Y_5, Z_5Y_6, Z_6Y_7, Z_9Y_{10}, Y_1Z_{10}, Z_1Y_2\}$ \\ \hline
12  & 3   & $\{Z_3Y_4, Z_4Y_5, Z_5Y_6, Z_7Y_8, Z_8Y_9, Z_9Y_{10}, Z_{11}Y_{12}, Y_1Z_{12}, Z_1Y_2\}$ \\ \hline
16  & 4   & $\{Z_3Y_4, Z_4Y_5, Z_5Y_6, Z_7Y_8, Z_8Y_9, Z_9Y_{10}, Z_{11}Y_{12}, Z_{12}Y_{13}, Z_{13}Y_{14}, Z_{15}Y_{16}, Y_1Z_{16}, Z_1Y_2\}$ \\ \hline
\end{tabular}
\end{table}

Division of the qubits into equally sized subsystems of $n/S$ qubits and this choice of Pauli-strings ensures $V$ to act along and between borders of these subsystems to capture longer-range correlations. For example, consider $n = 8$ and the measurement of $X_3$. According to Eq. \ref{eq:VdaggMVspecificproperties} this will result in,
\begin{align}
    V^\dagger X_3 V = \overline{P}^\dagger_3\overline{P}^\dagger_2 \overline{P}^\dagger_1 X_3 \overline{P}_1 \overline{P}_2 \overline{P}_3,
\end{align}
where $\overline{P}_k = \cos (c_k) I +i\sin (c_k) P_k$ for $\{P_1, P_2, P_3\} = \{Z_3Y_4, Z_4Y_5, Z_5Y_6\}$. Thus correlations among qubits 3,4,5 and 6 are captured. On the other hand, a measurement of $X_8$ will result in,
\begin{align}
    V^\dagger X_3 V = \overline{P}^\dagger_6\overline{P}^\dagger_5 \overline{P}^\dagger_4 X_8 \overline{P}_4\overline{P}_5 \overline{P}_6,
\end{align}
and capture correlations among qubits 1, 2, 7 and 8. For the Hamiltonian in Eq. \ref{eq:IsingHamiltonian}, such form for $V$ guarantees at most $2^{2K} = 2^{2\cdot 3} = 64$ expectation values per Hamiltonian term, regardless of system size.

\subsection{Supervised learning}
\label{subsec:MethodSupervisedLearning}
In this section, we will introduce the dataset and the structure of the model utilized for classification of handwritten digits.
\subsubsection{Dataset}
\label{subsec:dataset}
We will perform supervised learning on the digit dataset \cite{misc_pen-based_recognition_of_handwritten_digits_81}, which is the classification of handwritten digits from 0 to 9. Each sample is an 8 by 8 pixel image, which will require a 64 qubit quantum computer when encoding each pixel into its respective qubit. We will, similarly to \cite{Marshall2023highdimensional}, reduce this to a binary classification task of the numbers 3 and 6. The task will be performed on a simulation of an 8 qubit quantum computer and $\vec{x}^{(s)}$ in Eq. \ref{eq:QPCHypothesisClassSupervisedLearning} will thus contain the 8 pixels of the $s$'th column of the image. The pixels will be scaled to be in the range $[0, \pi]$ to serve as rotational angles in the parameterized ansatzes. The total number of samples in the dataset is 364, and these samples will be divided into 182 samples for training the model, 91 samples for validation and 91 samples for testing. 

\subsubsection{Subsystem unitaries}
\label{subsec:Ansatz}
The eight-qubit ansatzes corresponding to the unitary operators $U_s$ in Eq. \ref{eq:QPCHypothesisClassSupervisedLearning} is shown in Figs. \ref{fig:Ansatz} and \ref{fig:AnsatzComponents}.

\begin{figure}[H]

\Qcircuit @C=1em @R=1em {
& \multigate{3}{U(\vec{x}^{(s)})} & \multigate{3}{V(\vec{\theta}^{(s)}_1)} & \multigate{3}{W} & \multigate{3}{U(\vec{x}^{(s)})} & \multigate{3}{V(\vec{\theta}^{(s)}_2)} & \multigate{3}{W} & \multigate{3}{U(\vec{x}^{(s)})} & \multigate{3}{W} & \qw \\
& \ghost{U(\vec{x}^{(s)})} & \ghost{V(\vec{\theta}^{(s)}_1)} & \ghost{W} & \ghost{U(\vec{x}^{(s)})} & \ghost{V(\vec{\theta}^{(s)}_2)} & \ghost{W} & \ghost{U(\vec{x}^{(s)})} & \ghost{W} & \qw \\
\vdots & \pureghost{U(\vec{x}^{(s)})} & \pureghost{V(\vec{\theta}^{(s)}_1)} & \pureghost{W} & \pureghost{U(\vec{x}^{(s)})} & \pureghost{V(\vec{\theta}^{(s)}_2)} & \pureghost{W} & \pureghost{U(\vec{x}^{(s)})} & \pureghost{W}&   \\
& \ghost{U(\vec{x}^{(s)})} & \ghost{V(\vec{\theta}^{(s)}_1)} & \ghost{W} & \ghost{U(\vec{x}^{(s)})} & \ghost{V(\vec{\theta}^{(s)}_2)} & \ghost{W} & \ghost{U(\vec{x}^{(s)})} & \ghost{W} & \qw \\
}

\caption{The ansatz employed for the digits dataset. We utilize a data re-uploading scheme \cite{PerezSalinas2020datareuploading} with alternating layers of single-qubit rotations and entangling gates. The different components are explicitly written out in figure \ref{fig:AnsatzComponents}.}
\label{fig:Ansatz}
\end{figure}
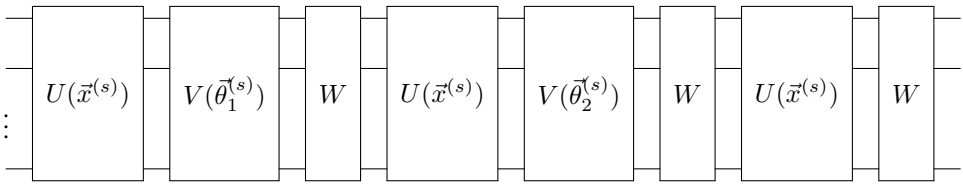

\begin{figure}[H]
  \centering
  \fbox{
  \begin{subfigure}{0.4\textwidth} 
    \centering
    
    \Qcircuit @C=1em @R=1em {
    & \gate{R_y(x^{(s)}_1)} & \qw \\
    & \gate{R_y(x^{(s)}_2)} & \qw \\
    & \vdots & \\
    & \gate{R_y(x^{(s)}_8)} & \qw \\
    }  
    \caption{ $U(\vec{x}^{(s)})$ }
  \end{subfigure}
  }
  \fbox{
  \begin{subfigure}{0.4\textwidth} 
    \centering
    \Qcircuit @C=1em @R=1em {
& \gate{R_y(\theta^{(s)}_1)} & \qw \\
& \gate{R_y(\theta^{(s)}_2)} & \qw \\
& \vdots & \\
& \gate{R_y(\theta^{(s)}_8)} & \qw \\
}
    \caption{$V(\vec{\theta}^{(s)})$}
  \end{subfigure}
  }
\fbox{
  \begin{subfigure}{0.3\textwidth} 
    \centering
    \Qcircuit @C=0.7em @R=0.7em {
& \ctrl{1} & \qw & \qw & \qw & \qw & \qw & \qw & \qw & \targ & \qw \\
& \targ & \ctrl{1} & \qw & \qw & \qw & \qw & \qw & \qw & \qw& \qw\\
& \qw & \targ & \ctrl{1} & \qw & \qw & \qw & \qw & \qw & \qw& \qw \\
& \qw & \qw & \targ & \ctrl{1} & \qw & \qw & \qw & \qw & \qw & \qw \\
& \qw & \qw & \qw & \targ & \ctrl{1} & \qw & \qw & \qw & \qw& \qw \\
& \qw & \qw & \qw & \qw & \targ & \ctrl{1} & \qw & \qw & \qw& \qw \\
& \qw & \qw & \qw & \qw & \qw & \targ & \ctrl{1} & \qw & \qw& \qw \\
& \qw& \qw& \qw& \qw& \qw& \qw& \targ & \qw & \ctrl{-7} & \qw& \qw\\
}
    \caption{$W$}
  \end{subfigure}
  }
  \caption{Figure showing the different components of the circuit in figure \ref{fig:Ansatz}.}
  \label{fig:AnsatzComponents}
\end{figure}
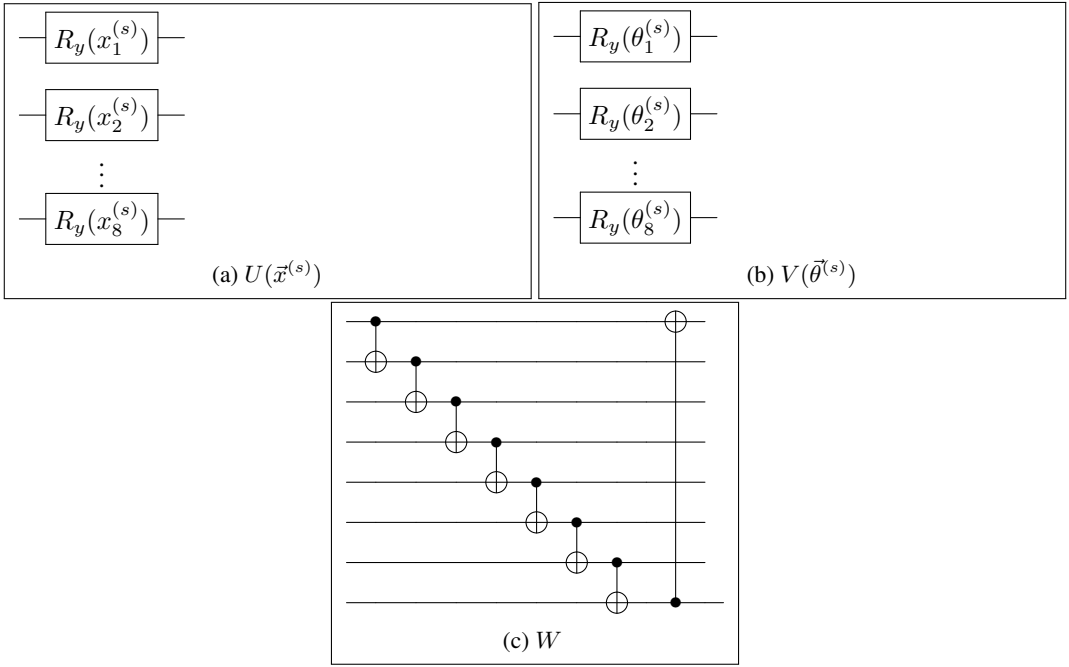
For the ansatzes, we apply a data re-uploading scheme to increase the expressiveness of our PQC. This is done by applying our feature encoding, $U(\vec{x}^{(s)})$ (see Fig. \ref{fig:AnsatzComponents} (a)), at several occasions in the ansatzes in Fig. \ref{fig:Ansatz}. It has been shown that the use this technique with only a single qubit can construct a universal quantum classifier \cite{PerezSalinas2020datareuploading}. The application of $V(\vec{\theta}^{(s)})$ (see Fig. \ref{fig:AnsatzComponents} (b)) encodes the tunable parameters of our model. This is done at two occasions with two different sets of parameters, $\vec{\theta}^{(s)}_1$ and $\vec{\theta}^{(s)}_2$, in order to further increase the expressiveness of the model. Since we are dealing with 8 subspaces, the total number of paramters for our ansatzes is $8 \cdot 16 = 128$. The applications of CNOT gates, $W$ (see Fig. \ref{fig:AnsatzComponents} (c)), is done to generate entanglement.

\subsubsection{Observable}
\label{subsec:Observable}
The observable for our task will alternate between measuring correlations between the first qubits in subsets $s$ and $s+1$, and correlations between the final qubits in subsets $s+1$ and $s+2$. That is
\begin{equation}
    \label{eq:DigitsMeasurementOperator}
    D^{(s,s+1)} = 
    \begin{cases}
        \sum_p d_p^{(s)}\sigma^{(s)}_{1,p} \sigma^{(s+1)}_{1,p},& \text{if $s$ is odd,}\\
        \sum_p d_p^{(s)}\sigma^{(s)}_{8,p} \sigma^{(s+1)}_{8,p},         & \text{if $s$ is even.}
    \end{cases}
\end{equation}
where $s \in \{1,2,\dots,,7 \}$ and $\sigma^{(s)}_{j,p}$ are one of the operators the Pauli basis, acting on the $j$'th qubit in subset $s$. We will restrict the coefficients $d_p^{(s)}$ to real scalars since our model outputs are real. If $n_b$ operators in the Pauli-basis are utilized in the construction of this observable, we require a total of $7\cdot 2n_b$ measurements per dataset sample and a total of $7\cdot 2^{n_b}$ coefficients to calculate the model output.

\section{Results}
\label{sec:Results}
We will now present the results for the approximation of the ground state energy of the 1D transverse-field Ising Hamiltonian, as well as for the classification of handwritten digits. We refer the reader to the Methods section \ref{sec:Method} for further details on how the models were configured.

\subsection{Ground state energy}
As detailed in section \ref{subsec:MethodGroundState},
the quantum system we will test out the method on is the 1D transverse-field Ising model with periodic boundary conditions, given by Eq. \ref{eq:IsingHamiltonian}. We aim to approximate the ground state energy for $n=4, 6, 8, 10, 12, 16$ spins and we define the global unitary operator $V$ according to Eq. \ref{eq:Vexpression}. The Pauli-strings $P_i$ utilized for the construction of $V$, along with the number of subsystems is shown in Tab. \ref{tab:Configuration}.

\begin{figure}[H]
    \centering
    \includegraphics[width=0.7\linewidth]{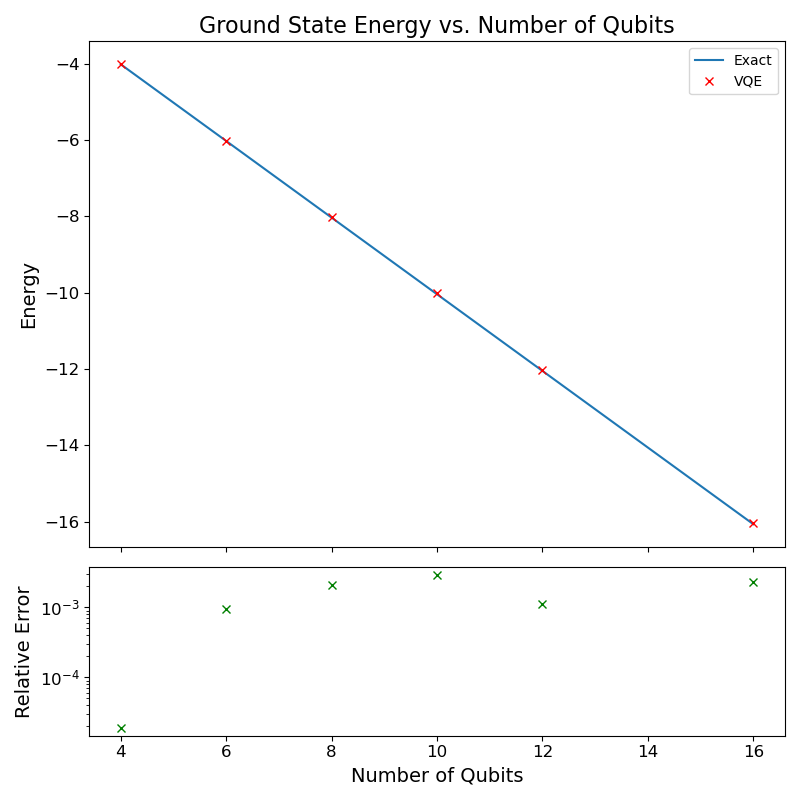}
    \caption{Exact ground state energy of the 1D transverse-field Ising model is compared with our approximation for $n$ number of spins/qubits. The top plot compares the achieved energies, while the bottom plot shows the relative error \(( E_{\text{VQE}}-E_{0})/E_{0}\) between the ground state energy $E_0$ and our approximation $E_G$.}
    \label{fig:GroundStateEnergy}
\end{figure}

The parameters of the PQC, $\vec{\theta}_s$, as well as the coefficients of $V$, that is $c_i$, are randomly initialized. The energy is then minimized by calculating the energy gradients in Eqs. \ref{eq:VQEGradientWrtRotationalParams} and \ref{eq:VQEGradientWrtCoefficients}, and utilizing gradient descent with the Adam optimizer\cite{adamoptimizer}. This is done with a learning rate of $0.1$ and $200$ training steps for every system. The results are compared with the exact diagonalization of the Hamiltonian in Eq. \ref{eq:IsingHamiltonian}, as shown in Fig. \ref{fig:GroundStateEnergy}.

The results of Fig. \ref{fig:GroundStateEnergy} demonstrates the capability to approximate the ground state energy with a relative error within the order of $0.1\%$ for all the tested numbers of qubits.

\subsection{Supervised learning}
Following the model structure in section \ref{subsec:MethodSupervisedLearning}, 
we trained two different instances of our model. First we utilized the full Pauli-basis, $\{I, \sigma_x, \sigma_y, \sigma_z \}$ in the construction of the observables in eq. \ref{eq:DigitsMeasurementOperator}. Since each of these observables only act on two subsets and one qubit in each, one model output corresponds to $56$ expectation values and we require $112$ coefficients. Since we have eight subsystems with 16 trainable rotational angles each, we have a total of $240$ trainable parameters in this model. For the second instance of the model, we utilized a truncated basis consisting of only the operators $\{I, \sigma_x \}$. One model output corresponds to a total of $28$ expectation values and we require $28$ coefficients. This leads to a total of $156$ trainable parameters. For both of these models, the mean squared error (MSE) was chosen as the loss function and the models were trained for $400$ epochs using the Adam optimizer with learning rate $0.1$. One epoch is defined as a single complete iteration over the entire training dataset. 

\begin{figure*}[ht!]
    \centering
    \begin{subfigure}[t]{0.49\textwidth}
        \centering
        \includegraphics[scale=0.55]{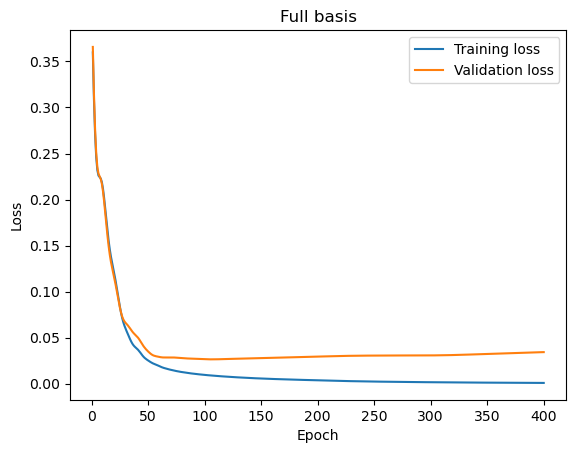}
        \caption{}
    \end{subfigure}%
    ~ 
    \begin{subfigure}[t]{0.49\textwidth}
        \centering
        \includegraphics[scale=0.55]{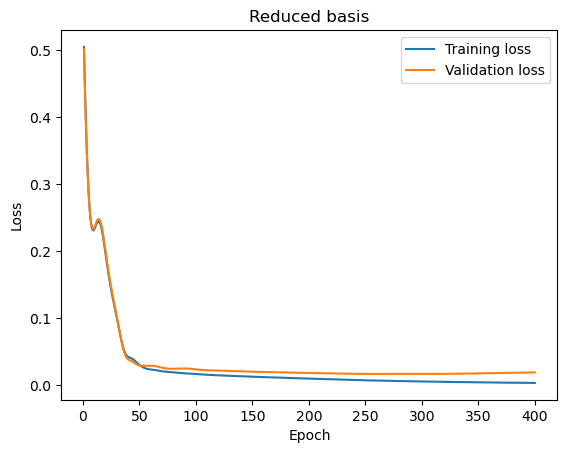}
        \caption{}
    \end{subfigure}
    \caption{MSE loss as a function of the number of epochs for the training and validation data for two different models. For both models, the final parameters corresponds to the parameters that minimized the validation loss. Figure (a) shows the loss when utilizing the full basis, $\{I, \sigma_x, \sigma_y, \sigma_z \}$ to construct the observables in eq. \ref{eq:DigitsMeasurementOperator}. For this case, the validation loss was minimized at epoch 105. Figure (b) shows the loss for when the reduced basis, $\{I, \sigma_x \}$ was utilized to construct the observables. The validation loss was minimized at epoch 250.}
    \label{fig:DigitResults}
\end{figure*}

Figure \ref{fig:DigitResults} shows the training and validation MSE as a function of the epoch for both models.
For both models, we see a rapid decrease of the training loss before seemingly converging towards a stable value. The validation loss also decreases similarly to the training loss in the early training stages, suggesting that the model generalizes to unseen data. Eventually, the validation loss starts to increase for both models, suggesting that overfitting might be occuring at some state. The final models were thus chosen from the epoch with the smallest MSE on the validation data. The resulting models were then applied on the remaining test data. The results are shown in table \ref{tab:AccMSE}.

\begin{table}[H]
\caption{The MSE and accuracy for the model utilizing the full basis (FB) of Pauli operators and the reduced basis (RB) consisting of $\sigma_a \in \{I, \sigma_x\}$.}
\centering
\begin{tabular}{|l|l|l|}
\hline
           & \begin{tabular}[c]{@{}l@{}}Accuracy (\%)\\ FB/RB\end{tabular} & \begin{tabular}[c]{@{}l@{}}MSE\\ FB/RB\end{tabular} \\ \hline
Training   & 100.0 / 100.0                                         & 0.0090 / 0.0076                                \\ \hline
Validation & 98.9 / 100.0                                          & 0.0265 / 0.0171                              \\ \hline
Testing    & 97.8 / 100.0                                          & 0.0304 / 0.0250                            \\ \hline
\end{tabular}
\label{tab:AccMSE}
\end{table}

Interestingly, the model utilizing the reduced basis is generalizing to unseen data better than the model with the full basis. This might be due to the basis reduction acting as a form of regularization, preventing overfitting. However, only one instance of each model have been trained, so these results might be coincidental. 

\begin{figure}[H]
    \centering
    \includegraphics[width=0.7\linewidth]{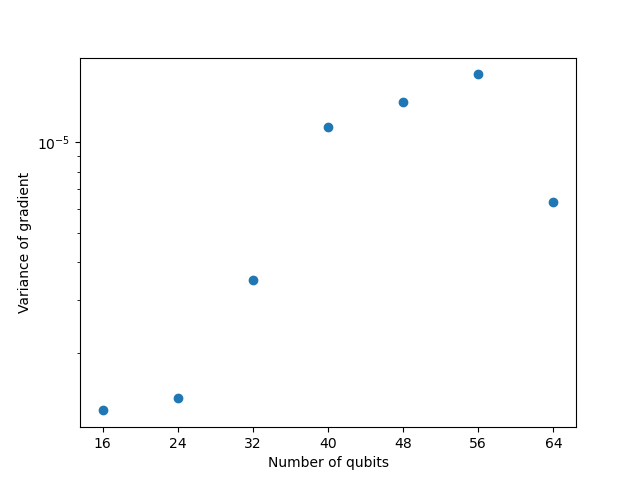}
    \caption{Variance of the gradient as a function of the number of qubits. The experiment starts with two subsystems, each containing 8 qubits, and additional subsystems of 8 qubits are incrementally added. For each configuration, 20 random parameter initializations are performed, and the gradient with respect to the first parameter in the first subsystem is sampled. The gradient is gathered from measurements of two-local Pauli-$X$ strings between pairs of subsystems.}
    \label{fig:vargrad}
\end{figure}

To illustrate that barren plateaus can be mitigated with this method, we start out with two subsystems, each of 8 qubits, and successively add one and one subsystem of 8 qubits until we have 64 qubits. For each of these configurations, we randomly initialize the model parameters and measure every possible two-local Pauli-string composed of Pauli-$X$ operators between two different subsystems, that is $\sum_{s<s'}\sum_{i j} d_{ijss'}X^{(s)}_i X^{(s')}_j$. This process is repeated for 20 random parameter initializations, where we sample the gradient with respect to the first parameter of the first subsystem. Based on the discussions in Section \ref{sec:barrenplateaus}, we anticipate that the variance of the gradient will not decrease exponentially as the number of subsystems increases. Figure \ref{fig:vargrad} illustrates the variance of the gradient as a function of the total number of qubits.

We observe from Fig. \ref{fig:vargrad} that when measuring two-local Pauli-$X$ strings between subsystems, the model does not suffer from exponentially vanishing gradients. This suggests that the chosen measurement strategy helps mitigate the barren plateau problem, often associated with deep quantum circuits. We suspect that the fluctuation in the variance is due to different parameter initializations.

\section{Conclusion}
\label{sec:Conclusion}

In conclusion, our study presents an approach for evaluating the expectation values of specific quantum states by decomposing the problem into smaller components to be evaluated on smaller quantum computers. The problem uses well-known techniques such as the measurement of expectation values of Pauli-strings and the parameter-shift rule\cite{ParameterShiftRule}, and can be tailored to mitigate the problem of barren plateaus. The approach can be utilized in the context of ground state energy approximation and also for quantum machine learning, which was demonstrated by applying the method to the 1D transverse-field Ising model with periodic boundary conditions, as well as for the digit dataset. For the ground state approximation, we achieved a relative error within the order of 0.1\% for all the tested system sizes of $n=2, 4, 6, 8, 10, 12, 16$ spins. For the digit dataset, the classification between the digits 3 and 6 showed that our model was able to generalize to unseen data with $100\%$ accuracy.

Looking ahead, we would like to apply the method to diverse quantum systems and datasets to assess its generalizability. Additionally, we would like to investigate the performance of our method in the presence of noise akin to currently available quantum devices. While the aforementioned studies are crucial for evaluating the potential for practical applications, this paper highlights the promise of our method and the potential in using small quantum devices for high-dimensional ground state approximation and machine learning tasks.

\section{Acknowledgements}
We would like to thank Morten Hjorth-Jensen for sharing his knowledge and engaging in helpful discussion.

\bibliographystyle{unsrtnat}
\bibliography{qmlqcp}  

\appendix
\section{Finding $E[B_s^{(i)} B_s^{(j)}]$ and $E[(B_s^{(i)})^2]$.}
\label{app:Introduction}
Consider a finite set $\{W_y \}_{y \in Y}$ with size $|Y|$, where $W_y$ are elements of $U(d)$, that is the unitary group of degree $d$. If $W_y$ forms unitary $t$-designs with $t\geq 2$, and $A,B,C,D$ are arbitrary linear operators, we have the following relation \cite{Costfunctiondependentbarreninshallowquantumcircuits},
\begin{align}
    \frac{1}{|Y|}\sum_{y \in Y} \text{Tr}[W_y A W_y^\dagger B]\text{Tr}[W_y C W_y^\dagger D]&=\int d\mu (U) \text{Tr}[U A U^\dagger B] \text{Tr}[U C U^\dagger D] \notag \\
    &= \frac{\text{Tr}[A]\text{Tr}[B]\text{Tr}[C]\text{Tr}[D] + \text{Tr}[AC]\text{Tr}[BD] }{d^2-1} \notag \\
    &- \frac{\text{Tr}[AC]\text{Tr}[B]\text{Tr}[D] + \text{Tr}[A]\text{Tr}[C]\text{Tr}[BD]}{d(d^2-1)},
    \label{eq:FindingVarianceTrace}
\end{align}
where $d$ is the dimensionality of the system and the integration is done over $U(d)$. In \cite{VanishingGradients}, it is argued that deep random circuits have the characteristics of 2-designs. Considering this for our ansatzes, $B_s^{(i)}$, we can use the relation in Eq. \ref{eq:FindingVarianceTrace} to calculate $E[B_s^{(i)} B_s^{(j)}]$ and $E[(B_s^{(i)})^2]$.

\subsection{$E[B_s^{(i)} B_s^{(j)}]$}
\label{app:BiBj}
Equation \ref{eq:FindingVarianceTrace} applied to $E[B_s^{(i)} B_s^{(j)}]$ gives
\begin{align}
    E[B_s^{(i)} B_s^{(j)}] =\int d\mu (U_s) \text{Tr}[U_s \rho U_s^\dagger W_s^{(i)}] \text{Tr}[U_s \rho U_s^\dagger W_s^{(j)}] \notag \\
    = \frac{\text{Tr}[\rho]^2 \text{Tr}[W_s^{(i)}]\text{Tr}[W_s^{(j)}] + \text{Tr}[\rho^2]\text{Tr}[W_s^{(i)} W_s^{(j)}]}{d^2-1} \notag \\
    - \frac{\text{Tr}[\rho^2] \text{Tr}[W_s^{(i)}]\text{Tr}[W_s^{(j)}] + \text{Tr}[\rho]^2\text{Tr}[W_s^{(i)} W_s^{(j)}]}{d(d^2 -1)},
\end{align}
where $d = 2^{n_s}$.
First off, $W^{(k)}_s$ are tensor-products of Pauli matrices and we have that $\text{Tr}[\bigotimes_i A_i] = \prod_i \text{Tr}[A_i]$. Since the Pauli matrices $X$, $Y$ and $Z$ have trace zero, this implies that any non-identity Pauli-string will have $\text{Tr}[W_s^{(i)}] = 0$. Since $W^{(i)}_s$ and $W^{(j)}_s$ are different Pauli-strings, at most one of them can be the identity-string. Thus $\text{Tr}[W_s^{(i)}]\text{Tr}[W_s^{(j)}] = 0$ for all $i\neq j$. Also, the products $W^{(i)}_s W^{(j)}_s \neq I$ since $W^{(i)}_s \neq  W^{(j)}_s $. This means that $\text{Tr}[W_s^{(i)} W_s^{(j)}] = 0$, and consequently,
\begin{align}
    \prod_{s\neq q} E[B_s^{(i)} B_s^{(j)}] = 0.
\end{align}

\subsection{$E[(B_s^{(i)})^2]$}
\label{app:Bsquared}
Equation \ref{eq:FindingVarianceTrace} applied to $E[(B_s^{(i)})^2]$ gives
\begin{align}
    \int d\mu (U_s) \text{Tr}[U_s \rho U_s^\dagger W_s^{(i)}] \text{Tr}[U_s \rho U_s^\dagger W_s^{(i)}] \notag \\
    = \frac{\text{Tr}[\rho]^2 \text{Tr}[W_s^{(i)}]^2 + \text{Tr}[\rho^2]\text{Tr}[(W_s^{(i)})^2]}{d^2-1} \notag \\
    - \frac{\text{Tr}[\rho^2] \text{Tr}[W_s^{(i)}]^2 + \text{Tr}[\rho]^2\text{Tr}[(W_s^{(i)})^2]}{d(d^2 -1)},
\end{align}
where $d = 2^{n_s}$. Assuming $\rho$ is a pure state such that $\text{Tr}[\rho] = 1$ and $\text{Tr}[\rho^2] = 1$, and rearranging gives
\begin{align}
    \int d\mu (U_s) \text{Tr}[U_s \rho U_s^\dagger W_s^{(i)}] \text{Tr}[U_s \rho U_s^\dagger W_s^{(i)}] \notag \\
    = (d-1)\frac{\text{Tr}[W_s^{(i)}]^2 + \text{Tr}[(W_s^{(i)})^2]}{d(d^2-1)}. \notag
\end{align}
Since $W^{(i)}_s$ is a tensor-product of Pauli matrices, $(W^{(i)}_s)^2 = I$, and thus $\text{Tr}[(W^{(i)}_s)^2] = d$. Further, we have that $\text{Tr}[\bigotimes_i A_i] = \prod_i \text{Tr}[A_i]$. Since the Pauli matrices $X$, $Y$ and $Z$ have trace zero, this implies that any non-identity Pauli string will have $\text{Tr}[W^{(i)}_s]^2 = 0$, while for identity Pauli-strings, $W^{(i)}_s = I$, we have $\text{Tr}[W^{(i)}_s]^2 = d^2$. We consider first the case when $W^{(i)}_s = I$. We then get
\begin{align}
    &\int d\mu (U_s) \text{Tr}[U_s \rho U_s^\dagger W_s^{(i)}] \text{Tr}[U_s \rho U_s^\dagger W_s^{(i)}] \notag \\
    &= \frac{(d-1)(d^2 + d)}{d(d^2-1)} \notag \\
    &= \frac{(d-1)(d+1)}{(d^2-1)} \notag \\
    &= 1,
\end{align}
which is expected from measurements of the identity operator on normalized states.
When $W^{(i)}_s \neq I$ we get
\begin{align}
    &\int d\mu (U_s) \text{Tr}[U_s \rho U_s^\dagger W_s^{(i)}] \text{Tr}[U_s \rho U_s^\dagger W_s^{(i)}] \notag \\
    &= \frac{d(d-1)}{d(d^2-1)} \notag \\
    &= \frac{d-1}{d^2-1} \notag \\
    &= \frac{d-1}{(d-1)(d+1)} \notag \\
    &= \frac{1}{d+1}, \notag \\
\end{align}
For $d=2^{n_s} = 2^{\beta \log_2(n)} = n^\beta$ we thus have,
\begin{equation}
E\left[(B_s^{(i)})^2\right] = 
\begin{cases} 
1 & \text{if } W_s^{(i)} = I \\ 
\frac{1}{n^\beta+1} & \text{otherwise}.
\end{cases}
\end{equation}
If $\bigotimes_{s\neq q}^S W^{(i)}_s$ acts non-trivially only on $K$ of the subsystems, it follows that
\begin{equation}
    \prod_{s\neq q}^S E[(B_s^{(i)})^2] = \frac{1}{(n^\beta + 1)^K},
\end{equation}
which shows that terms of $D = \sum_p d_p \bigotimes_{s=1}^S W_s^p$ which are local to a small subset of $K$ of the subsystems can have the appropriate scaling for the variance of the gradients.

\end{document}